\documentclass[a4paper,11pt]{article}
\usepackage{pos}
\usepackage{braket}
\usepackage{amsmath}
\usepackage{amssymb}
\usepackage{multicol}
\usepackage{slashed}
\usepackage{mathtools}
\usepackage{float}
\usepackage{dsfont}
\usepackage{caption}
\usepackage{subcaption}
\usepackage{lipsum}

\usepackage[utf8]{inputenc}\DeclareUnicodeCharacter{2212}{-}
\usepackage[normalem]{ulem}

\title{Confining Strings and Glueballs in $\mathbb{Z}_N$ Gauge Theories}

\newcommand{\be}{\begin{equation}}
\newcommand{\ee}{\end{equation}}
\newcommand{\bea}{\begin{eqnarray}}
\newcommand{\eea}{\end{eqnarray}}

\author[a]{Andreas Athenodorou}
\author[b]{Sergei Dubovsky}
\author*[b]{Conghuan Luo}
\author[c,d]{Michael~Teper}

\affiliation[a]{Computation-based Science and Technology Research Center, The Cyprus Institute, Cyprus}
\affiliation[b]{Center for Cosmology and Particle Physics, Department of Physics, New York University, New York, NY, 10003, USA}
\affiliation[c]{Rudolf Peierls Centre for Theoretical Physics, University of Oxford, Parks Road, Oxford OX1 3PU, UK}
\affiliation[d]{All Souls College, University of Oxford, High Street, Oxford OX1 4AL, UK}

\abstract{ Effective string theory has shown its universal power in the prediction of the spectrum of low-lying excited states of confining strings. Here  we  study confining flux tubes in $\mathbb{Z}_N$ gauge theories. For the $N=2$ theory, which corresponds to the 3d  Ising gauge model, we compute the spectrum of low-lying excitations of confining strings and show that it agrees with the universal Nambu--Goto predictions except for an additional  massive scalar resonance. This resonance, however, turns out to be a bulk glueball mixing with the flux tube excitations rather than a genuine string worldsheet state. In general $\mathbb{Z}_N$ gauge theories (dual to clock spin models), we observe a continuous phase transition for $N \geq 4$, while for $N > 5$ it is governed by the $O(2)$ universality class. The critical behavior of the string tension and mass gap is verified to be described by a dangerously irrelevant operator. At large $N$ the glueball spectrum is expected to approach the spectrum of U(1) gauge theory, which is confirmed by our lattice data.}

\FullConference{%
  The 40th International Symposium on Lattice Field Theory (Lattice2023),\\
  31 July - 4 August, 2023 \\
  Fermilab, Illinois, USA 
}

\begin{document}
\maketitle

\section{Introduction}
\label{sec:introduction}

Understanding the color confinement in Quantum Chromodynamics (QCD) has been a long-standing problem ever since early 1970s. Simply put, this puzzle can be phrased as the question of how does a non-Abelian gauge theory, that describes  colored quarks and gluons at high energies, produce a spectrum of colorless hadrons at long distances? The concept of confinement becomes precise, for theories like  pure glue $SU(N_c)$ Yang--Mills  model which enjoy the $\mathbb{Z}_N$ one-form electric ``center" symmetry. The order parameter for the center symmetry is the vacuum expectation value (VEV) of Polyakov loops, which translates into the area law for the Wilson loops $\langle W(C) \rangle \sim e^{-\text{Area(C)}/\ell_s^2}$. This implies the existence of stable long strings of a finite tension \footnote{See, e.g. \cite{Gaiotto:2014kfa,McGreevy:2022oyu} for reviews on higher-form symmetries. } $T=\ell_s^{-2}$. This notion can be generalized to any theory with a one-form symmetry, such as 3D $\mathbb{Z}_N$ gauge theories, which are the main focus of this talk. Studying the behavior of such confining strings may shed light on our understanding of confinement. 

Recently, there has been a revival of research on QCD and confinement in terms of effective string theory. This is stimulated in part by the development of high precision Monte-Carlo computations of confining string spectrum for different theories \cite{Athenodorou:2010cs,Athenodorou:2011rx}. 
The spectrum of strings wrapped around a spatial circle of circumference $R$ can be approximated using the derivative expansion based on the effective string theory. 
Assuming that translational Goldstone  modes are the only massless  degree of freedom  on the worldsheet, the Nambu--Goto  action  fixes the form of the string energy levels up to  subleading non-universal  corrections which  start at the    $O(\ell_s^6/R^7)$ order in the derivative expansion~\cite{Aharony:2013ipa,Dubovsky:2012sh}. 

On the other hand, for a  while it was  observed that the following expression  
\begin{equation}
    E_{\text{GGRT}}(N_L, N_R)=\sqrt{\frac{4 \pi^2(N_L - N_R)^2}{R^2}+\frac{R^2}{\ell_s^4}+\frac{4 \pi}{\ell_s^2}\left(N_L + N_R-\frac{D-2}{12}\right)} \,,
    \label{GGRT}
\end{equation}
 (dubbed the Goddard–Goldstone–Rebbi–Thorn (GGRT) formula) provides  an  exceptionally   accurate approximation  for the energies   of many low-lying string states, even when the  strings are relatively short and the $\ell_s/R$ expansion fails. Originally this formula was obtained by  the light-cone quantization of the Nambu-Goto action~\cite{goddard1973quantum,arvis1983exactqq}. However, this quantization is known to be incompatible with the spacetime Poincar\'e symmetry. This puzzle has been  resolved  when it was realized that (\ref{GGRT}) arises as the leading   order answer  in the Thermodynamic Bethe Ansatz (TBA)  method of calculating the string spectrum~\cite{Dubovsky:2012wk,Dubovsky:2013gi,Dubovsky:2014fma}, which has better convergence properties than the straightforward $\ell_s/R$ expansion. 
 
 The resulting improved theoretical control provides an opportunity  to search for novel non-universal string excitations associated with the states which strongly deviate from  (\ref{GGRT}).  For example, the 4d Yang-Mills confining string is found to support an extra massive pseudo-scalar mode ~\cite{Dubovsky:2013gi,Dubovsky:2014fma}.  
 It is natural to ask whether massive resonances are also present  on the worldsheet of $\mathbb{Z}_N$ strings. Here we present the results for the Ising model, which corresponds to $N=2$. As we will see, a new subtlety arises for  $\mathbb{Z}_N$ strings. Namely, bulk glueballs may mix with the genuine string excitations and mimic the presence of a resonance.
  In the Yang-Mills theory such a mixing is suppressed at large $N_c$. No such suppression is present in the $\mathbb{Z}_N$ case so  we need to find a way to distinguish bulk glueballs from genuine worldsheet states. 

Another motivation for studying $\mathbb{Z}_N$ gauge theories is to understand how the  $U(1)$ gauge theory is recovered in the $N \to \infty$ limit, where confinement comes out as a result  of monopole condensation \cite{polyakov1975compact,polyakov1977quark}. In particular, what is the mechanism for the confinement in $\mathbb{Z}_N$ gauge theories, and why deconfining transitions are only present  for $\mathbb{Z}_N$ gauge theories, unlike in the $U(1)$ case?

The rest of this note is organized as follows.  We sketch our methodology of lattice simulations in section~\ref{sec:lattice}. Then we review the results of 3d Ising strings in section~\ref{sec:Ising} based on \cite{Athenodorou:2022pmz}. In section~\ref{sec:zn} we summarize some new results for $\mathbb{Z}_N$ gauge theories. The details of the data and analysis will be presented in a forthcoming publication \cite{zngaugetheory}.

\section{Lattice setup}
\label{sec:lattice}

We use the Wilson lattice gauge theory action
\begin{equation}
    S = \beta \sum_{\rm plaq} \{1- \operatorname{Re} 
    U_{\rm plaq}\} \,, \quad U_{\rm plaq}(n,\mu,\nu) = U_{\mu}(n) \cdot U_{\nu}(n+\hat{\mu}) \cdot U_{\mu}^{\dagger}(n+\hat{\nu}) \cdot U_{\nu}^{\dagger}(n)\;,
\end{equation}
where the link fields take  values $U_{\mu}(n) = e^{\frac{2\pi i k}{N}}, \quad k \in \mathbb{Z}/N\mathbb{Z}$. This is called the (vector) $\mathbb{Z}_N$ gauge theory. The simulation is performed using the Metropolis algorithm. We typically make over $10^5$ measurements for each lattice system and coupling. 

We compute the low-lying spectrum of confining strings and glueballs from two point functions of operators that represents the physical objects we are interested in, located at different times. We use Polyakov operators that wind around the spatial dimension once to represent confining strings, and contractable Wilson loops to represent glueballs. The excited states are extracted by constructing a large set of operators, and using variational methods to find out linear combinations of operators that have large overlap with  the low-lying states. Smearing and blocking methods are applied. Further details of the simulation can be found  in \cite{Athenodorou:2010cs,Athenodorou:2022pmz,Teper:1998te}. 

Operators are characterized by a set of quantum numbers determined by their symmetry properties. Operators from different sectors have zero two point functions. For winding strings, the quantum numbers include the longitudinal momentum $p = 2\pi q/R$ (the transverse momentum is fixed to zero in our simulations), the transverse parity $P_{\perp}$, that flips the sign of the transverse spatial coordinate, and the longitudinal parity $P_{\parallel}$ that flips the sign of the longitudinal coordinate\footnote{This latter symmetry corresponds to the $CP$ symmetry of the underlying gauge theory, because it  reverses the flux direction.}. Note that $P_{\parallel}$ is not a good quantum number for states with $q \neq 0$. Correspondingly, we label the string states either by $(P_{\perp} P_{\parallel})$ or by $q \neq 0$ and $(P_{\perp})$. 
On the other hand, glueball states are characterized by their spin $J$, parity $P$ and charge conjugation $C$ (not present for Ising), and we label the corresponding sectors as $|J|^{PC}$.

\section{3d Ising strings}
\label{sec:Ising}

The Kramers-Wannier duality states that the 3d $\mathbb{Z}_2$ lattice gauge theory (LGT) has the same partition function as the Ising model with no external magnetic field. This duality exchanges the low temperature phase of one model to the high temperature phase of another, and vice versa \cite{savit1980duality}. This implies that the critical behavior of the second order deconfining transition in the $\mathbb{Z}_2$ gauge theory is governed by 3D Ising universality class. 

\begin{figure}[hbt]
\centering
\scalebox{1}{\includegraphics[width=0.6\textwidth]{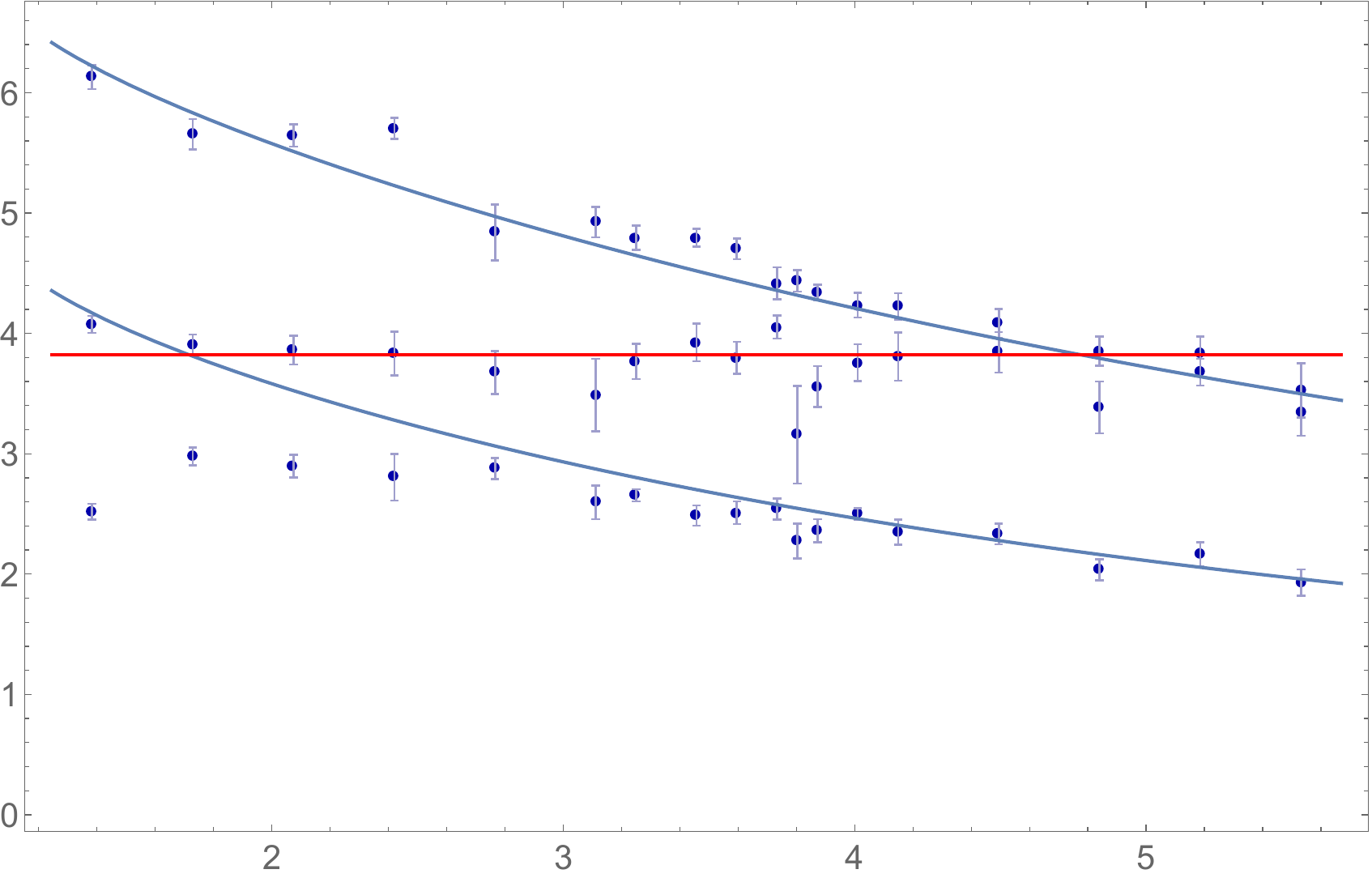}\put(-280,160){\scriptsize $\Delta E\ell_s$}\put(0,0){\scriptsize $R/\ell_s$}}
\caption{$q=0$ excited states in the $(++)$ parity sector as a function of string length.  Blue curves are the $(1,1)$ and $(2,2)$ GGRT spectrum. The horizontal red line is the fitted mass for the massive resonance state.  }
\label{plotparityplus}
\end{figure}

We focus on the confining phase and take the continuum limit of the gauge theory by sending $\beta$  to $\beta_c = 0.7614133(22)$ \cite{blote1999cluster}, where the string tension and the mass gap are small in the lattice unit.
We find that the lowest-lying states of wound flux tubes agree with the GGRT prediction~\eqref{GGRT}, with some notable exceptions. Specifically, we observe a massive resonance in the $(++)$ sector with $m_{res}\ell_s = 3.825(50)$, shown in Figure~\ref{plotparityplus}. A similar state shows up as a boosted massive resonance in the $q=1;(+)$ sector. 

To determine whether this state is a genuine massive worldsheet resonance  or a bulk excitation (glueball), we note that a bulk excitation can have transverse momentum relative to the string. So if this state is a glueball, in principle we should also be able to observe a series of scattering states of a glueball and a string, with quantized relative momentum $p_{\perp} = 2\pi q_{\perp}/l_{\perp}$, but the operators we use may have poor overlap onto them. To probe these state we added  ``multitrace" operators of the following form
\begin{equation}
\phi_{\text {scattering }}=\sum_{n, m=1}^{l_{\perp} / a} \phi_P(y+n a) \phi_G(y+m a) e^{\frac{2 \pi i q_{\perp}(n-m) a}{l_{\perp}}} \,,
\end{equation}
where $\phi_P$ represents a string state, and $\phi_G$ is a glueball. Our initial simulation, which suggested the presence of a resonance, was performed without these operators and at fixed $l_{\perp}\approx 4.8\ell_s$.
In order to identify scattering states, recall that $q_{\perp}\neq 0$ scattering states are expected to exhibit a  strong transverse volume dependence corresponding to the dispersion relation,
\begin{equation}
    E=\sqrt{m_{\text {flux }}^2+p_{\perp}^2}+\sqrt{m_{\text {glue }}^2+p_{\perp}^2}, \quad p_{\perp}=2 \pi q_{\perp} / \ell_{\perp} \,.
\end{equation}

\begin{figure}[hbt]
    \scalebox{1}{\includegraphics[width=0.6\textwidth]{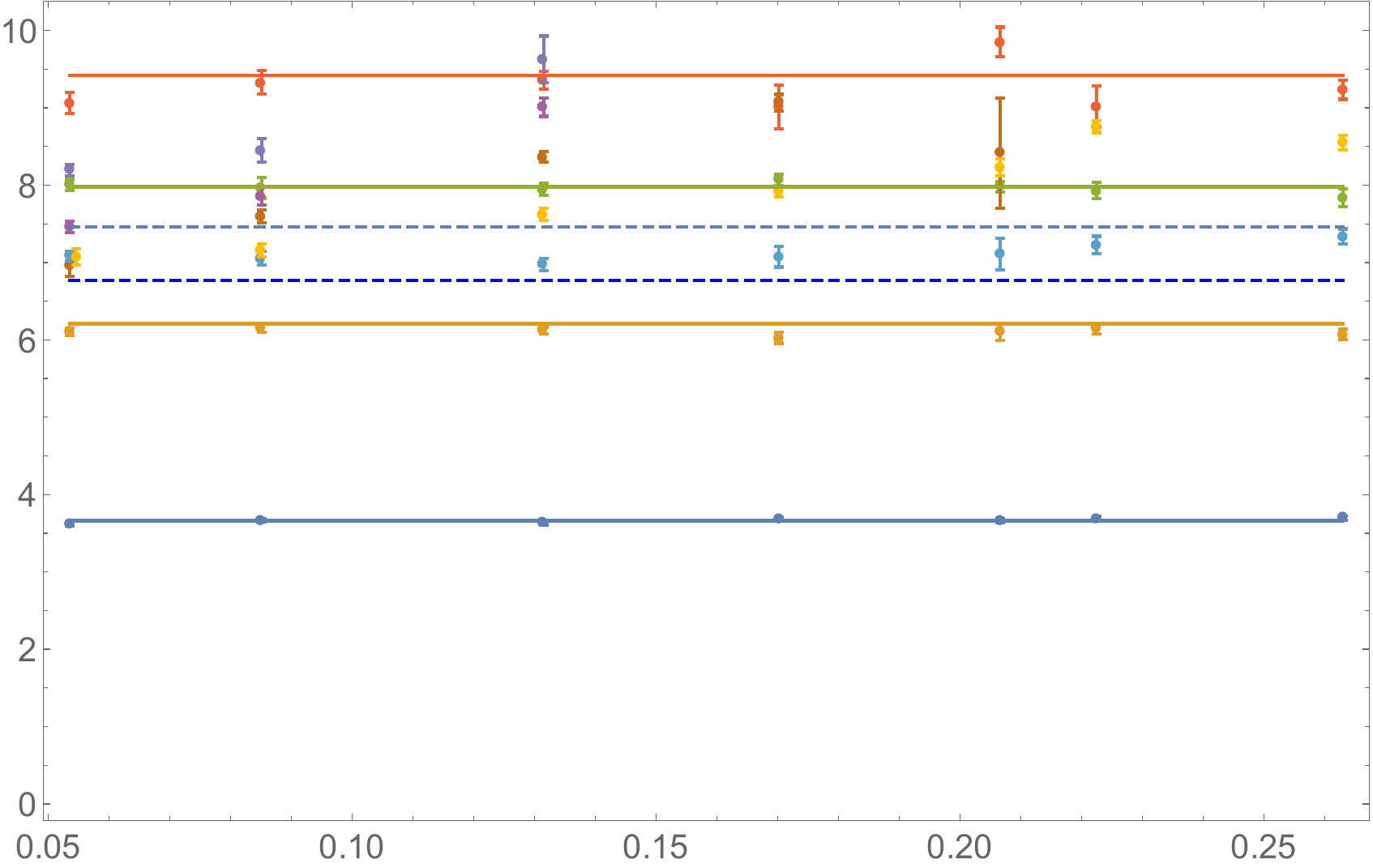}\put(-280,160){\scriptsize $E \ell_s$}\put(0,0){\scriptsize $\ell_s/l_{\perp}$}}
\centering
\caption{Spectrum in the $(++)$ sector at $R=55a=3.80l_s$ as a function of the inverse transverse volume with scattering operators. The GGRT spectrum is represented by horizontal solid lines of different colors 
starting from $N=\tilde{N}=0$. 
The energy of the absolute ground state plus the glueball mass is shown by the lower dashed blue line. The absolute ground state plus the resonance mass is shown by the upper dashed blue line.}
\label{plot_ppvolume_5glue}
\end{figure}

We present the transverse volume dependence of the $(++)$ sector after including scattering operators up to $q_{\perp}=4$ in Figure~\ref{plot_ppvolume_5glue}. 

We observe all four scattering states with $q_{\perp} \neq 0$ as expected, but no new $q_\perp=0$ state in  addition to the  ``resonant" state. The conclusion is that the resonance is in fact a $q_\perp=0$  glueball, rather than a genuine massive resonance on the string.

\section{3d $\mathbb{Z}_N$ gauge theories}
\label{sec:zn}

The Kramers-Wannier duality can be extended to the duality between generalized $\mathbb{Z}_N$ clock spin model and the (vector) $\mathbb{Z}_N$ LGT
\cite{Borisenko:2012nf,ukawa1980dual}. It has been suggested that clock spin models with $N \geq 5$ exhibit a second order phase transition governed by the $O(2)$ universality class \cite{hove2003criticality,Borisenko:2013xna}.
Observables sensitive to the  $U(1)\to\mathbb{Z}_N$ breaking typically exhibit an intricate scaling behavior associated with a dangerously irrelevant deformation \cite{oshikawa2000ordered}. The duality dictates then that there should be a deconfining transition in the gauge theory, which is also governed by the $O(2)$ universality class. However, so far relatively little is  known about the confining mechanism of the $\mathbb{Z}_N$ LGT and how it approaches the
$U(1)$ limit as $N \to \infty$.
We aim to fill this gap here.

\subsection{(De)confinement mechanism}

The confining behavior is associated with the area law for the vev of the Wilson loop, $\langle W(C) \rangle$. In the $U(1)$ case the partition function  can be rewritten as a Coulomb gas of magnetic monopoles \cite{polyakov1975compact,polyakov1977quark}. The area law arises as a result of 
the (magnetic) Debye screening of the magnetic field sourced by the external current (the Wilson loop).
This behavior persists at all values of the coupling constant $\beta$, which controls the monopole number density. 

One expects the monopoles to be confined in the $\mathbb{Z}_N$ case. One way to argue for this is to realize the $\mathbb{Z}_N$ theory as arising from the spontaneous symmetry breaking $U(1)\to \mathbb{Z}_N$, where the monopole confinement arises as a consequence of the conventional (not the dual one) Meissner effect. 

Directly in the $\mathbb{Z}_N$ language one observes that the non-vanishing magnetic flux through a plaquette in the $\mathbb{Z}_N$ model cannot be smaller than $2\pi/N$. Hence, an individual $U(1)$ monopole can be approximate by a $\mathbb{Z}_N$ gauge field configuration only up to a distance of order 
$l_T \simeq \sqrt{N/4\pi} a$.
At longer distances a spherically symmetric $U(1)$  monopole turns into a $\mathbb{Z}_N$ junction of $N$ magnetic strings carrying the minimal magnetic flux. To determine the tension of such a string note that a  magnetic string loop can be generated from the vacuum configuration by setting all links intersecting a surface bounded by the loop to $e^{\frac{2\pi i }{N}}$. The action of this configuration is proportional to the length $l$ of the loop
\begin{equation}
\Delta S_s(l) = \beta [1 - \cos(2\pi/N)] l\;.
\end{equation}

These considerations suggest the following dynamics of  $\mathbb{Z}_N$ models at small values of $\beta$. First, there is a "$U(1)$ region" at the smallest values of $\beta$,
$0<\beta\lesssim\beta_N$, such that the typical monopole separation in the $U(1)$ model is smaller than $l_T$. Here one expects the $U(1)$ and $\mathbb{Z}_N$ spectra to be practically indistinguishable, because in both cases the dynamics is dominated by $U(1)$-like monopole configurations. At larger values of $\beta$, $\beta_N\lesssim\beta<\beta_c$ the model is expected to exhibit $\mathbb{Z}_N$ confinement, where instead of the monopole plasma the Wilson line is screened by a network of  $\mathbb{Z}_N$ junctions connected by magnetic strings.

The critical value $\beta_c$, where the theory deconfines can be estimated by noting that the density of states  of a large magnetic string loop
 grows exponentially with its length $l$, so that the corresponding contribution to the partition function can be estimated as
\begin{equation}
    Z_{\text{string}} \approx e^{\gamma l - \beta \left[1 - \cos\left( \frac{2\pi}{N} \right) \right] l} \,,
\end{equation}
where $\gamma$ is a constant. Its specific value is not important for the argument. We then expect that the string network disappears and the theory deconfines at 
\begin{equation}
\label{betac}
\beta_c = \frac{\gamma}{1 - \cos(2\pi/N)} \,.
\end{equation}
 This functonal form of the deconfining point agrees well with the lattice results of \cite{bhanot1980phase}, 
 which can be fitted by (\ref{betac}) withh $\gamma\approx 1.5$ (which also reproduces the known values of $\beta_c$ at $N\leq 4$).
 As $N \to \infty$, the deconfining point is pushed to infinity. 

To confirm the existence of the small $\beta$ $U(1)$ region we preformed lattice simulations which determine the mass spectrum and the string tension at the fixed relatively small $\beta$ and different $N$.
 The results are presented  in Table~\ref{table:recover_u1}, which indeed indicate that at $\beta=2.20$ the spectrum of $\mathbb{Z}_N$ models with $N\geq 8$ is indeed very close to the $U(1)$ spectrum. On the other hand, the $\mathbb{Z}_6$ theory at  $\beta=2.20$ appears to exhibit the $\mathbb{Z}_N$ confinement.
\begin{table}[htbp]
\centering
\begin{tabular}{c|cc|ccc}
\hline \hline \multicolumn{6}{c}{$Z(N)$ and $U(1)$ lightest masses and string tension } \\
\hline \hline group & $\beta$ & $L_s^2 L_t$ & $a M_{0^{++}}$ & $a M_{0^{--}}$ & $a \sqrt{\sigma} $ \\
\hline
$U(1)$ & 2.20 & $48^3$ & $0.5386(23)$ & $0.2691(14)$ & $0.16646(62)$ \\
$Z(100)$ & 2.20 & $48^3$ & $0.5320(23)$ & $0.2648(12)$ & $0.16683(50)$ \\
$Z(10)$ & 2.20 & $48^3$ & $0.5367(23)$ & $0.2673(17)$ & $0.16488(39)$ \\
$Z(8)$ & 2.20 & $48^3$ & $0.5267(92)$ & $0.2644(17)$ & $0.16469(76)$ \\
$Z(6)$ & 2.20 & $48^3$ & $0.451(9)$ & $0.2167(18)$ & $0.14252(51)$ \\
\hline
\end{tabular}
\caption{The recovery of $U(1)$ spectrum as $N$ gets larger. }
\label{table:recover_u1}
\end{table}

\subsection{Critical behavior}

Now we turn to the critical behavior of observables in $\mathbb{Z}_N$ gauge theories. In particular, we focus on the bulk mass gap and the string tension. The question is how to relate the physical quantities we measure to the well-known CFT data in the $O(2)$ CFT. One might expect the critical exponents to be determined by the dimensions of the relevant operators, but this model provides an example to this behavior, associated with dangerously irrelevant operators. 

Let us  present an renormalization group (RG) flow argument that prescribes the critical behavior. Consider the UV Gaussian fixed point ($u=g=\lambda_N=0$) in the complex scalar field theory,
\begin{equation}
    S = \int d^3 x \left[ |\partial_{\mu} \Phi|^2 + u |\Phi|^2 + g |\Phi|^4 + \lambda_N (\Phi^N + \bar{\Phi}^N) \right] \,.
\end{equation}
If one tunes $u=u_c=0$ and inspects a family of theories with different $g>0$, one finds the critical value $g_c$ such that the theory runs into the $O(2)$ IR fixed point. Perturbing this RG trajectory fixed point by introducing $u>0$, one triggers an RG flow which starts at the Gaussian UV fixed point, approaches the vicinity of the $O(2)$ fixed point and then deviates into the Nambu--Goldstone theory,
\begin{equation}
    S = \int d^3 x K (\partial_{\mu} \theta)^2 + c\lambda_N K^3 \cos N\theta \,.
    \label{NG}
\end{equation}
The latter describes the (pseudo)-Goldstone mode of the $O(2)$ spontaneous symmetry breaking. 
Here we introduced also the leading $O(2)$ breaking operator, which is generated if one perturbs the Gaussian theory by turning on $\lambda_N\neq 0$, with $c$ being a dimensionless constant.
Interestingly, this symmetry breaking term is irrelevant both at the Gaussian and the $O(2)$ fixed points, but becomes relevant at the NG fixed point, providing an example of a dangerously irrelevant behavior. 

It was proposed in \cite{oshikawa2000ordered}  that the confining $\mathbb{Z}_N$ theory near criticality is described exactly by this RG flow.
The critical behavior is then governed by the charge $N$ operator as well as the typical length scale $K$ in the $O(2)$ CFT,
\begin{equation}
    c\lambda_N \sim (T-T_c)^{\nu |y_N|}, \quad K \sim 1/\xi \sim (T-T_c)^{\nu} \,,
\end{equation}
where $\nu = 0.6718(1)$~\cite{Chester:2019ifh}, $|y_5| = 1.27(1)$, $|y_6|=2.55(6)$~\cite{shao2020monte}. As a consequence of the dangerously irrelevant behavior,  we obtain a very weakly coupled massive theory in the deep IR. This is supported by the lattice simulations, where $M_{0^{++}} \approx 2 M_{0^{--}}$ for $\mathbb{Z}_N$ gauge theories.

The scaling of the mass gap and the string tension is then determined by~\eqref{NG}, 
\begin{equation}
    m^2 = \lambda_N K^2 \sim (T - T_c)^{\nu (|y_N| + 2)}, \quad \sigma \propto Km \sim (T - T_c)^{\nu (\frac{|y_N|}{2} + 2)} \,.
\end{equation}
For $N=5$, we verified these scalings quite precisely,
\begin{equation}
    a M_{0^{--}}(\beta) = 0.94(13) (2.17961 - \beta)^{1.09(5)}, \quad a^2 \sigma(\beta) = 0.183(8) (2.17961 - \beta)^{1.725(20)} \,.
\end{equation}
For $N=6$, our results are less accurate and we are not able to approach sufficiently close to the critical point, because the mass gap is much smaller in lattice unit, and lattice simulations suffer from severe finite volume effects. Nevertheless, we still get a qualitative agreement (within $3\sigma$),
\begin{equation}
    a M_{0^{--}}(\beta) = 0.26(3) (3.0068 - \beta)^{1.90(12)}, \quad a^2 \sigma (\beta) = 0.0212(21) (3.0068 - \beta)^{2.00(11)}
\end{equation}

\section{Conclusion}

In this proceeding, we summarize our simulation results of the 3d Ising string spectrum, and introduce a method to probe scattering states of a glueball and a string, with which we have identified the only "resonance" state we have observed is a glueball mixed with the string, instead of a genuine massive resonance. We also explain the (De)confinement mechanism and the critical behaviors of 3d $\mathbb{Z}_N$ gauge theories and present simulational evidence supporting them.




\section*{Acknowledgements}
We thank John Donahue, Yifan Wang for fruitful discussions. AA has been financially
supported by the EuroCC2 project funded by the Deputy Ministry of Research, Innovation and
Digital Policy and the Cyprus Research and Innovation Foundation and the European High-
Performance Computing Joint Undertaking (JU) under grant agreement No 101101903. SD and MT
acknowledge support by the Simons Collaboration on Confinement and QCD Strings. SD is also supported in part by the NSF grant PHY-2210349, and by the BSF grant 2018068.

\bibliographystyle{JHEP}
\bibliography{biblio_NEW}
\end{document}